\documentclass[preprint2]{aastex}
\usepackage{graphicx}
\usepackage{txfonts}

\shorttitle{correlated optical and gamma emissions from GRB~081126}
\shortauthors{Klotz et al.}

\begin{document}


\title{
Observation of correlated optical and gamma emissions from GRB~081126
}

\author{A. Klotz}
\affil{Observatoire de Haute-Provence, F--04870 Saint Michel l'Observatoire, France}
\affil{CESR, Observatoire Midi-Pyr\'en\'ees, CNRS, Universit\'e de Toulouse, 
BP 4346, F--31028 Toulouse Cedex 04, France}
\email{alain.klotz@cesr.fr}

\author{B. Gendre}
\affil{LAM/Universit\'e de Provence/CNRS, 38 rue Joliot Curie, F--13388 Marseille Cedex 13, France}

\author{J.L. Atteia}
\affil{LATT, Observatoire Midi-Pyr\'en\'ees, CNRS, Universit\'e de Toulouse, 
14 Avenue E. Belin, F--31400 Toulouse, France}

\author{M. Bo\"er}
\affil{Observatoire de Haute-Provence, F--04870 Saint Michel l'Observatoire, France}

\and

\author{D.M. Coward \& A.C. Imerito}
\affil{School of Physics, University of Western Australia, M013, Crawley WA 6009, Australia}



\date{Received {\today} /Accepted }

\begin{abstract}
   We present an analysis of time-resolved optical emissions observed from the gamma-ray burst \objectname{GRB~081126} 
   during the prompt phase. The analysis employed time resolved photometry using optical data obtained by the TAROT 
   telescope, using BAT data from the {\it Swift} spacecraft
   and time resolved spectroscopy at high energies from the GBM instrument onboard the \emph{Fermi} spacecraft. 
   The optical emission of \objectname{GRB~081126} is found to be compatible with the second gamma emission pulse shifted 
   by a positive time-lag of 8.4$\pm$3.9\,sec. 
  This is the first well resolved observation of a time lag between optical and gamma emissions during a gamma--ray burst. 
  Our observations could potentially provide new constraints on the fireball model for gamma ray burst early emissions. 
  Furthermore, observations of time-lags between optical and gamma ray photons provides an exciting opportunity to constrain quantum gravity theories.
\end{abstract}


\maketitle

\section{Introduction}
Because of their extreme luminosity in $\gamma$-rays, gamma ray bursts (GRBs) are a unique probe to high energy 
regimes where exotic physics is likely to manifest. A fraction of GRBs have been associated with the collapse of 
massive stars via the association of supernova signatures observed with the fading GRB optical 
afterglow e.g. \citep{hjorth03,stan03}. The afterglow most likely originates from an external shock produced as the 
blast wave from the progenitor collides with the interstellar medium,
 causing it to slow down and lose energy. 
Fast moving telescopes linked to GCN notices \citep{Barthelmy1998} are able
to record the optical counterpart at the time when the
prompt $\gamma-$ray emission is still active. The first positive detection of such event 
was GRB~990123 \citep{akerlof99}. Some other successful detections have been achieved so far
\citep[e.g.][]{Rascusin08}.
Two general results have been seen: Either a bright optical emission, uncorrelated to the gamma-ray light curve,
occurred \citep[for 5 to 20\% of GRBs according to][]{klotz09}, or that a faint optical emission 
is correlated with the gamma-ray flares \citep[\objectname{GRB~050820A},][]{Vestrand2006}. 
In the former case, these bright optical flashes are often interpreted as the reverse shock signature \citep{Jin2007}.
\\
Time lags between X--ray and gamma-ray data are often observed
\citep[e.g.][]{Norris00}. However, this is rare between optical and $\gamma-$rays.
As an example, \cite{Tang06} estimated the most probable time lags for the light curves of \objectname{GRB~990123}
(5--7\,sec) and GRB~041219A (1--5\,sec). However, the optical data
have poor time sampling, putting doubts on these results. 
Moreover, no lag was noticed for GRB~041219 by \citet{Zheng2006}.
No lag was reported for GRB~050820A at a level of few seconds \citep{Vestrand2006}.
\\
In this letter, we present the measurements of the optical
emission observed by TAROT \citep{klotz08} during the prompt $\gamma-$ray activity of \objectname{GRB~081126}. 
We show evidence for a positive time lag between optical and $\gamma-$ray light curves.

\section{GRB~081126}
GRB~081126 (Swift BAT trigger 335647, with T$_0$=26th Nov. 2008, 21:34:10 UT)
light curve shows a small precursor starting at $\sim$T$_0-$30 s,
peaking at $\sim$T$_0-$18 s, and returning almost to zero at T$_0-$7 s \citep{Sato08}. The burst  
features two peaks, the first one at $\sim$T$_0+$1.5 s, reaching its maximum
at $\sim$T$_0+$7 sec. The second one peaks at $\sim$T$_0+$31.5 sec.
The duration of that burst is T$_{90} = 54\pm4$\,s (15-350 keV).
This event was also detected by Konus-Wind \citep{Golenetskii08} and the Fermi GBM 
\citep{Bhat08}.

The time-averaged spectrum of the first pulse from T$_0$ to T$_0+$11 s is
well fit by a Band function with E$_{peak}$ = 192 $\pm$ 74 keV,
alpha = -0.3 $\pm$ 0.4, and beta = -1.6 $\pm$ 0.1.
The second pulse from about T$_0+$20 s to T$_0+$40 s is also
well fit by a Band function with E$_{peak}$ = 162 $\pm$ 77 keV,
alpha = -0.3 $\pm$ 0.5, and beta = -1.6 $\pm$ 0.1. 
The fluence (8-1000 keV) in the two pulses are
(2.7 $\pm$ 0.8)$\times$10$^{-7}$ erg\,cm$^{-2}$ and (1.9 $\pm$ 0.8)$\times$10$^{-7}$ erg\,cm$^{-2}$
respectively.

XRT observation reported by \cite{Margutti08} started 65.7 seconds after
the BAT trigger, too late to gather X--ray information of the second peak. 
The XRT detected a characteristic afterglow emission of the burst. This afterglow was not detected by a quick visual
inspection of images taken by TAROT, started 20.6 s after the burst
\citep{gendre08}. However, \cite{Skvarc08} reported the optical light curve of the afterglow in R band 
using the 60\,cm of the Crni Vrh Observatory. Their observations
start at T$_0$+82s. They observe a slow rise in optical emission that peaks 200s 
after the trigger and then fades. This optical afterglow was also reported by 
\cite{Andreev08}, using the Z-600 telescope of Mt. Terskol observatory, 33 minutes
after the burst, and by UVOT \citep{Holland08} at
21$^h$34$^m$03.59$^s$ +48$^\circ$42'38.3" (J2000.0).
They report that the  detection in the U filter,
combined with the lack of detections in the UV filters, is consistent
with the afterglow having a redshift of approximately $2.4<z<3.8$ \citep{Holland08}.
Unfortunately, no other photometric observations were performed to improve this estimation.
From Konus-Wind data, we deduced a pseudo-redshift of $5.3 \pm 1.8$ using the method described in \cite{att03}.
\\
The Galactic latitude of the afterglow position is -2.29$^\circ$
and the corresponding extinction is E(B-V)=0.782 mag. according to 
\cite{Schlegel1998}. Assuming $R$=3.1, this gives A$_{V}$=2.6 and 
A$_{R}$=2.1 mag.
\\
\section{TAROT data}

\begin{figure}[htb]
\centering
\includegraphics[width=0.9\columnwidth]{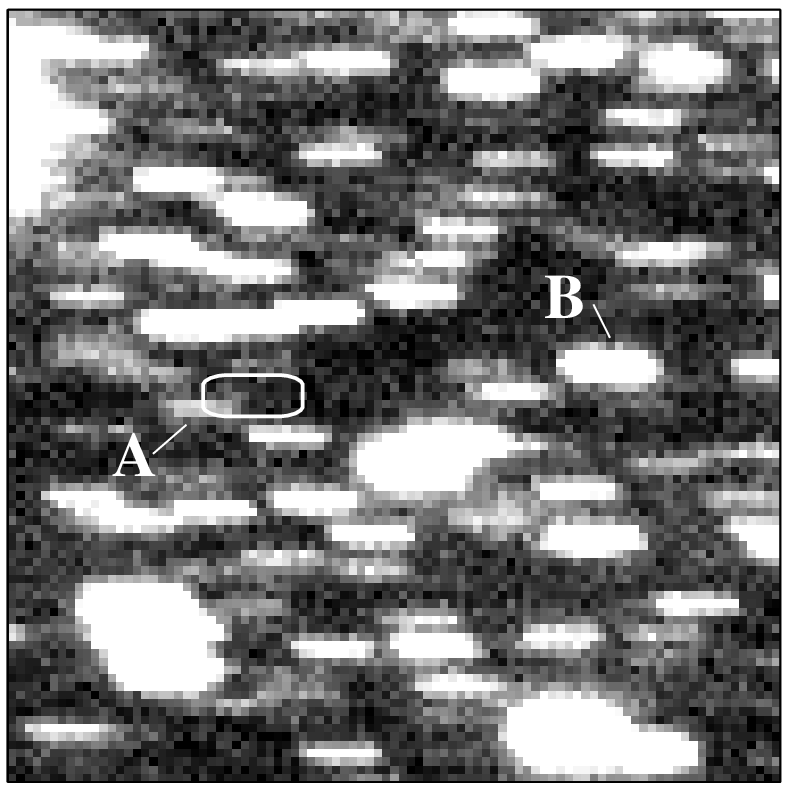}
\vspace*{5mm}
\includegraphics[width=0.9\columnwidth]{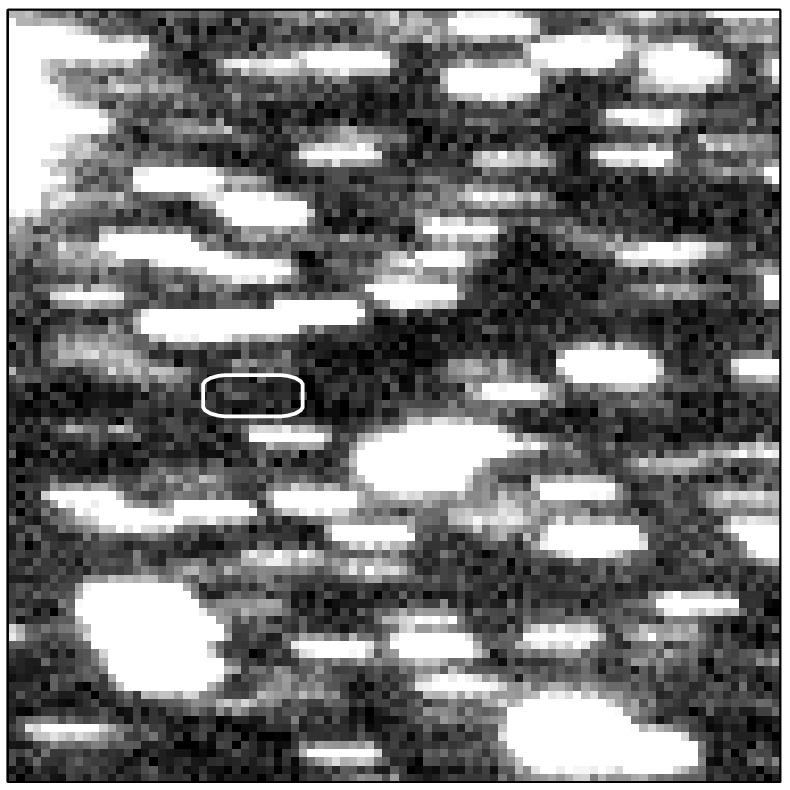}
\caption{Field of GRB~081126. Top: TAROT image taken between 21s and 81s after the GRB trigger.
The hour angle velocity was adapted to obtain stars as trails
of $\sim$9.2 pixel length during the 60s exposure.
The theoretical position of the GRB trail is indicated by the white box.
The star A covers partly the GRB trail.
(see text).
Bottom: After subtraction of star A using the model of star B, the trace of
the optical emission of the GRB appears in the box
The image size is 5 arcmin,
North is up, East left.
\label{trail}} 

\end{figure}

The first TAROT images were obtained at T$_0+$20.1s (duration 60\,sec)
with the tracking speed adapted to obtain a small trail 
of a few pixel length. This technique
is used in order to obtain temporal informations during
the exposure \citep[e.g.][]{Klotz2006}. The spatial sampling is 3.29\,arcsec/pix and the FWHM
of stars (in the perpendicular direction of the trail) is 2.05 pixels.
On the trailed image (see Figure~\ref{trail}), the flux of the afterglow is
affected by the proximity of NOMAD1 1387-0420537 (R=18.1) but
also by the end of the trail of NOMAD1 1387-0420579 (R=15.48 hereafter A). This last
star lies at 21 arcsec East and 7\,arcsec South from the GRB
position. As a matter of consequence, the trail of star A
(which spreads over 30\,arcsec) covers partially the beginning of the
trail of the GRB (Fig.~\ref{trail} top).\\

\begin{figure}[htb]
\centering{\includegraphics[width=1\columnwidth]{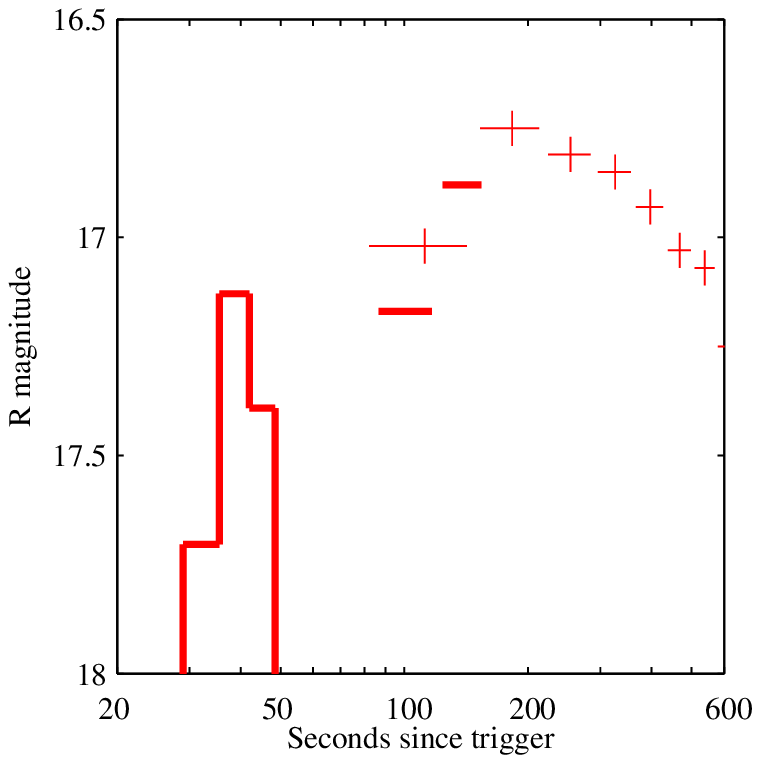}}
\caption{
Optical light curve of GRB~081126. TAROT optical data are thick bars and
observations from \cite{Skvarc08} are thin bars (2 sigma level).
There are data in the ranges 20--29 and 49--89\,sec, 
but with no detection at the limit of R=18.0.
} \label{tarot_crni}
\end{figure}

Knowing the position of the afterglow, we first subtracted the trail of the star A within the image. 
The star NOMAD1 1387-0420302 (R=13.17, hereafter B) is far enough to other neighbor
stars to be used as a trail template to model the star A.
We then subtracted this model from the image (using a correct scaling factor to take into account 
the difference of flux between the stars A and B). The result of the
subtraction shows clearly the presence of a dim optical emission 
(Fig.~\ref{trail} bottom).\\
\\
Successive images are 30\,s long exposures tracked on the diurnal motion. \cite{gendre08} published 
only upper limits using TAROT data because it was impossible to detect the optical counterpart so 
close to the star NOMAD1 1387-0420537 without careful subtraction. The images taken later 
by TAROT were employed to perform this subtraction. The technique successfully revealed the optical afterglow.
In Fig.~\ref{tarot_crni} we display the initial part of the TAROT light curve. We add data from
\cite{Skvarc08} showing that we can distinguish the early
emission that occurred during the gamma activity and the
afterglow that followed. A discussion of the afterglow emission process is beyond the scope of this 
paper and will be presented in Corsi et al. (2009, in preparation).

\section{Data analysis}

\begin{figure}[htb]
\centering{\includegraphics[width=1\columnwidth]{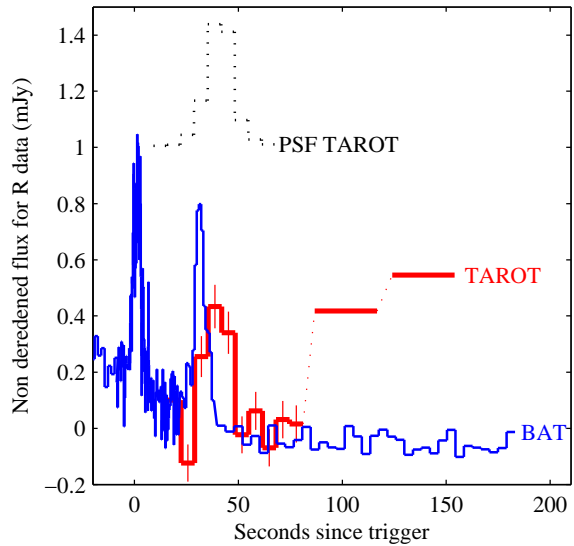}}
\caption{
Light curves of GRB~081126 measured by BAT and TAROT.
The dotted line labeled 'PSF-TAROT' stands for the spread
of a star equivalent to an instantaneous flash of 0s duration.
This figure appears in colors in the electronic version.
} \label{tarot_bat}
\end{figure}

From the trailed image, a horizontal profile corresponding to the 
predicted position of the afterglow gives directly the light curve.
We measured the temporal sampling of 6.5 sec/pixel using trails
of bright stars. The light curve of the afterglow in the trail presents a flare within 3 pixels 
(Fig. \ref{tarot_bat}), with a probability of $\sim10^{-8}$ to be spurious. The probability to 
observe a cosmic ray at that position is $3.6\times 10^{-6}$ (estimated from dark fields of the same night). 
We thus conclude that this flare is real and produced by the burst itself.

Such light curve profiles are affected by the 
Point spread function (PSF) of TAROT. In order to compare the optical and $\gamma-$ray light curves, 
we need to convolve the BAT signal by the TAROT PSF.
The PSF can be extracted as a vertical profile of a bright non saturated star
(seen as the doted curve in the Figure\,\ref{tarot_bat}).
We performed a symetrisation of the PSF shape to be compatible with
the hypothesis that the PSF shows no direction effect.
Once convolved with the TAROT PSF, the BAT signal corresponding to one peak is very similar to that of TAROT.

We note at the start of the trail a bright single pixel that could be associated with the end of 
an optical flare. However, this event is not significant enough to be used in our analysis. 
Nonetheless, it could be an optical flare related to the first $\gamma-$ray pulse. 
In the following analysis, we will consider this as a possibility, and thus that the optical flare is 
correlated to the second $\gamma$-ray pulse, without discarding the possibility that the optical 
flare is linked to the first $\gamma$-ray pulse. 

\begin{figure}[htb]
\centering{\includegraphics[width=1\columnwidth]{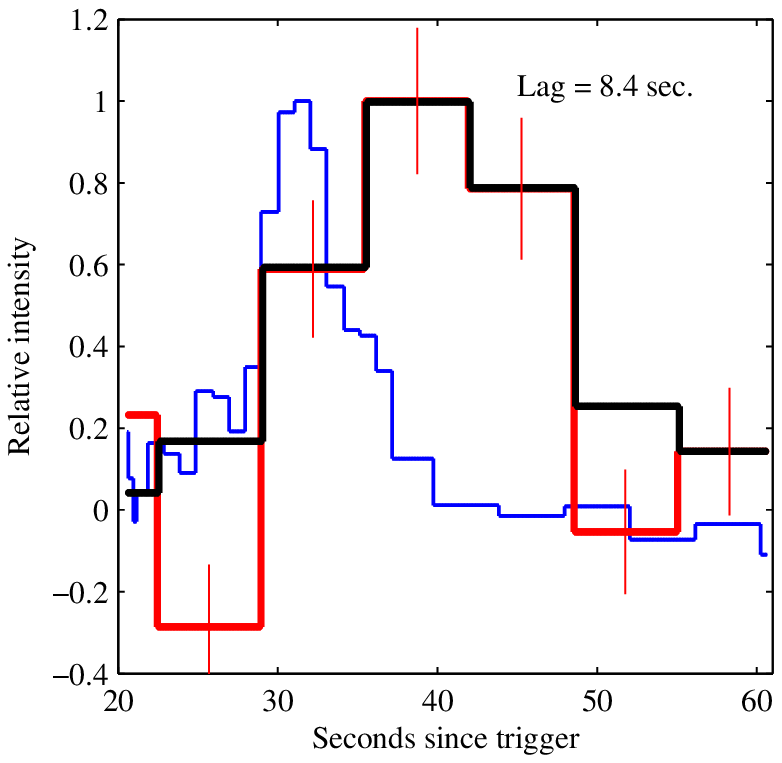}}
\caption{
The convolution of the peak of BAT light curve (in blue) by the PSF-TAROT
shifted by 8.4s (in black) compared to the TAROT data (in red).
This figure appears in colors in the electronic version.
} \label{tarot_lag}
\end{figure}

A $\chi^{2}_{\nu}$ fit between the optical flare and $\gamma-$ray pulse implies a temporal lag of $+8.4 \pm 3.9$ s 
(see Fig. \ref{tarot_lag}) at the 97\% confidence ($+38.4 \pm 3.9$ s if the optical flare 
is related to the first $\gamma$-ray pulse). 
This is strong evidence for a positive time-lag between the optical and high energy feature. 
We point out that the exposure time of  TAROT images has a better accuracy than 0.1s because 
we use a GPS card triggered by the opening of the shutter, and is not dependent on the computer 
internal clock variations \citep{laas2008}.

As the TAROT PSF is larger than the BAT second pulse, we also studied the influence of
the duration of the BAT pulse modelized by a Gaussian shape, letting free the width 
of the Gaussian within the fit. 
The best match of the modelized BAT pulse is a Gaussian
spread by sigma=4.0\,sec. The $\chi^{2}_{\nu}$ fit gives the same lag as for the actual BAT pulse
meaning that the profile shape of the pulse does not constrain the lag value.
The fit remains compatible for Gaussians with sigma lower than 9\,sec. This means the
optical pulse is compatible with a high energy pulse which could have a duration between 0 to 
9\,sec.

The flux of the optical peak observed by TAROT is 0.45 mJy. To be compared with the Fermi observations, 
this value must be corrected for two effects: i) the spread of the flux due to the PSF
profile, and ii) the large optical extinction in the R band. Correcting for all these effects, 
the optical flux is $\sim$6\,mJy at the peak. 
We used the Band model parameters obtained by the Fermi-GBM \citep{Bhat08} to compute the 
optical flux expected from the high energy band. We derived an expected optical flux of 
2.6$\times 10^{-10}$\,Jy, which is $\sim10^{-7}$ times the one observed.
Taking account for the uncertainties in the Bhat's alpha parameter
the extrapolated flux is always $\sim10^{-5}$ times the one observed.

\section{Discussion and conclusions}

The analysis of optical and gamma--ray light curves of GRB~081126 reveals: i) 
the width of the optical peak is the same as the gamma--ray peaks, ii)
the profile of the optical peak is consistent to the gamma--ray peaks 
after correcting for the different PSF, iii) the optical 
peak occurred 8.4$\pm$3.9\,s (or $+38.4 \pm 3.9$ s) later than the gamma peak. This is the first time-lag measured between
optical and gamma light curves of a GRB.  iv) the gamma--ray flux measured by GBM Fermi, extrapolated 
to optical energies is $\sim10^{-7}$ times smaller than the 
optical flux. 

These three results provide potentially new constraints on the theory of prompt GRB emissions.
For example, time lags between different energy photons are predicted by quantum gravity 
in the framework of string theory \citep[e.g.][]{Amelino98}.
However, in such a case optical photons should arrive
before gamma ones. As we observe the opposite, one can rule out this hypothesis
for the GRB~081126's optical lag.
Gamma-ray photons comptonization on cold electrons could explain the profile 
of the optical flare. However, this cannot explain the positive lag observed.

Within the internal shock framework, this temporal lag implies that optical photons were emitted 
after the $\gamma-$ray ones. However, it is surprising that the flux {\it increases} so dramatically 
during this process. 
This is not well understood 
in the standard model for the inelastic internal shock and our results provide new tools for refining the standard model.

%
\begin{acknowledgements}
B. Gendre acknowledges support from {\it Centre National
d'Etudes Spatiales} (CNES).
The TAROT telescope has been funded by the {\it Centre National
de la Recherche Scientifique} (CNRS), {\it Institut National des
Sciences de l'Univers} (INSU) and the Carlsberg Fundation. It has
been built with the support of the {\it Division Technique} of
INSU. We thank the technical staff contributing to the TAROT project,
G. Buchholtz, J. Eysseric, M. Merzougui, C. Pollas, P. Richaud and Y. Richaud.

\end{acknowledgements}


\begin{thebibliography}{}

\bibitem[Amelino-Camelia et al.(1998)]{Amelino98} Amelino-Camelia, G., Ellis, J., Mavromatos, N.E., et al., 1998, Nature, 393, 763
\bibitem[Andreev et al.(2008)]{Andreev08} Andreev M., Sergeev A., Pozanenko, A., 2008, GCNC 8558
\bibitem[Akerlof et al.(1999)]{akerlof99} Akerlof, C., et al., 1999, Nature, 398, 400
\bibitem[Atteia(2003)]{att03} Atteia, J.L., 2003, A\&A 407, L1
\bibitem[Barthelmy(1998)]{Barthelmy1998} Barthelmy, S., in AIP Conf. Proc. 428, Gamma--Ray bursts, ed. C. Meegan \& R. Preece (Berlin: Springer), 99
\bibitem[Bhat(2008)]{Bhat08} Bhat, P.N., van der Horst, A.J., 2008, GCNC 8589
\bibitem[Gendre et al.(2008)]{gendre08} Gendre, B., Klotz, A., Atteia, J.L., et al., 2008, GCNC 8555
\bibitem[Golenetskii et al.(2008)]{Golenetskii08} Golenetskii, S., Aptekar, R., Mazets, E., et al., 2008, GCNC 8562

\bibitem[Hjorth et al.(2003)]{hjorth03}Hjorth J., et al., 2003, Nature, 423, 847

\bibitem[Holland et al.(2008)]{Holland08} Holland, S.T., Evans, P.A., Marshall, F.E., et al., 2008, GCNC 8564
\bibitem[Jin \& Fan(2007)]{Jin2007} Jin, Z.P. \& Fan, Y.Z., 2007, MNRAS 378, 1043
\bibitem[Klotz et al.(2006)]{Klotz2006} Klotz, A., Gendre, B., Stratta, G., et al., 2006, A\&A, 451, L39
\bibitem[Klotz et al.(2008)]{klotz08} Klotz, A., Bo\"er, M., Eysseric, J., et al., 2008, PASP, 120, 1298
\bibitem[Klotz et al.(2009)]{klotz09} Klotz, A., Bo\"er, M., Atteia, J.L., et al., 2009, AJ, in press
\bibitem[Laas-Bourez et al.(2008)]{laas2008} Laas-Bourez, M., Bo\"er, M., Blanchet, G., et al. 2008. Cospar 37, 1672
\bibitem[Margutti et al.(2008)]{Margutti08} Margutti, R., Beardmore, A.P. , Brown, P.J., 2008, CGNC 8554
\bibitem[Norris et al.(2000)]{Norris00} Norris, J.P., Marani, G.F., Bonnell, J.T., 2000, ApJ 534, 248
\bibitem[Rascusin et al.(2008)]{Rascusin08} Rascusin, J.L., Karpov, S.V., Sokolowski, M., et al., 2008, Nature, 455, 183
\bibitem[Sato et al.(2008)]{Sato08} Sato, G., Barthelmy, S.D., Baumgartner, W.H. , 2008, GCNC 8557
\bibitem[Skvarc \& Mikuz(2008)]{Skvarc08} Skvarc, J., Mikuz, H., 2008, GCNC 8569
\bibitem[Schlegel et al.(1998)]{Schlegel1998} Schlegel, D.J., Finkbeiner, D.P., Davis, M., 1998, ApJ, 500, 525
\bibitem[Stanek et al.(2003)]{stan03}Stanek K.Z, Matheson, T., Garnavich, P.M., et al. 2003, ApJ, 591, L17
\bibitem[Tang \& Zhang(2006)]{Tang06} Tang, S.M., Zhang, S.N., 2006, A\&A 456, 141
\bibitem[Vestrand et al.(2006)]{Vestrand2006} Vestrand W.T, Wozniak, P.R., Wren, J.A., et al., 2006, Nature 442, 172
\bibitem[Zheng et al.(2006)]{Zheng2006} Zheng Z., Ye L., Yong-Heng Z., 2006, ApJ 646, L25
\end{thebibliography}
\end{document}